\documentclass[preprint,aps]{revtex4}
\usepackage{graphicx}% Include figure files
\usepackage{dcolumn}% Align table columns on decimal point
\usepackage{bm}% bold math
\newcommand{\ef}{$E_F$}

\newcommand{\ccs}{CsCr$_3$Sb$_5$}
\newcommand{\cvs}{CsV$_3$Sb$_5$}
\DeclareUnicodeCharacter{2212}{-}
\begin{document}

\title{Exotic Surface Stripe Orders in Correlated Kagome Metal \ccs}% 
%\title{Exotic Surface Density-Wave Stripes in the Hund's Kagome Superconductor \ccs}% Force line breaks with \\
\author{Yunxing Li$^{1,*}$, Peigen Li$^{1,2,*}$, Taimin Miao$^{3,*}$, Rui Xu$^{1}$, Yongqing Cai$^{4}$, Neng Cai$^{3}$, Bo Liang$^{3}$,\\ Han Gao$^{5}$, Hanbo Xiao$^{5}$, Yongzhen Jiang$^{5}$, Jiefeng Cao$^{6}$, Fangyuan Zhu$^{6}$,  Hongkun Wang$^{1,2}$,\\ Jincheng Xie$^{1,2}$, Jingcheng Li$^{1}$, Zhongkai Liu$^{5}$, Chaoyu Chen$^{7}$, Yunwei Zhang$^{1}$, X. J. Zhou$^{3,\dagger}$,\\ Dingyong Zhong$^{1,2,\dagger}$, Huichao Wang$^{1,\dagger}$, Jianwei Huang$^{1,\dagger}$, Donghui Guo$^{1,\dagger}$}

\affiliation{
\\$^{1}$Guangdong Provincial Key Laboratory of Magnetoelectric Physics and Devices, Center for Neutron Science and Technology, School of Physics, Sun Yat-sen University, Guangzhou 510275, China
\\$^{2}$State Key Laboratory of Optoelectronic Materials and Technologies, Sun Yat-sen University, Guangzhou, 510275 China
\\$^{3}$Institute of Physics, Chinese Academy of Sciences, Beijing, 100190 China
\\$^{4}$School of Physics, Dalian University of Technology, Dalian, 116024 China
\\$^{5}$School of Physical Science and Technology, ShanghaiTech Laboratory for Topological Physics, ShanghaiTech University, Shanghai, 201210 China
\\$^{6}$Shanghai Synchrotron Radiation Facility, Shanghai Advanced Research Institute, Chinese Academy of Sciences, Shanghai, 201204 China
\\$^{7}$Songshan Lake Materials Laboratory, Dongguan, 523808 China
\\$^{*}$ These authors contributed equally
\\$^{\dagger}$ X. J. Zhou: xjzhou@iphy.ac.cn;~~~~~~~~~~~~ \\Dingyong Zhong: dyzhong@mail.sysu.edu.cn; \\Huichao Wang: wanghch26@mail.sysu.edu.cn; \\Donghui Guo: guodonghui@mail.sysu.edu.cn; \\Jianwei Huang: huangjw269@mail.sysu.edu.cn
}

\date{\today}% It is always \today, today,

\begin{abstract}
The newly discovered kagome superconductor \ccs~exhibits distinct features with flat bands and unique magnetism, providing a compelling platform for exploring novel quantum states of correlated electron systems. Emergent charge order in this material is a key for understanding unconventional superconductivity, but it remains unexplored at the atomic scale and the underlying physics is elusive. Here, we identify and unreported stripe orders on the surface which are distinct from the bulk and investigate the underlying bulk electronic properties using a combination of scanning tunneling microscopy (STM), angle-resolved photoemission spectroscopy (ARPES) and density functional theory (DFT) calculations. %The 4$a_0$ $\times$ $\sqrt{3}a_0$ charge order is indicated by a clear contrast inversion pattern upon bias reversal. 
Specifically, a mixture of $2a_0\times a_0$ and $3a_0\times a_0$ stripe order is found on Cs-terminated surface while 4$a_0$ $\times$ $\sqrt{3}a_0$ stripe order is found on the Sb-terminated surface. The electronic spectra exhibit strongly correlated features resembling that of high temperature superconductors, with kagome flat bands lying about 330 meV above \ef, suggesting that the electron correlations arise from Coulomb interactions and Hund's coupling. Moreover, a distinct electron–boson coupling mode is observed at approximately 100 meV. These findings provide new insights into the interplay between surface and bulk charge orders in this strongly correlated kagome system.

\end{abstract}

\maketitle

%\tableofcontents

\section{\label{sec:level1}Introduction}
Quantum materials with a kagome lattice serve as a promising platform for exploring diverse novel phenomena. Its unique geometrically frustrated lattice structure induces phase-dependent annihilation phenomena in nearest-neighbor transitions for bosons or fermions, giving rise to a series of intriguing electronic structures including linearly dispersive Dirac node, van Hove singularities (vHS), and flat bands~\cite{Yin2022, Wang2023, Neupert2022, Wilson2024, Jiang2023, Zhang2025}. In 2019, the discovery of the vanadium-based kagome superconductor AV$_3$Sb$_5$ (A = K, Rb, Cs) sparked wide interest and intensive research in condensed matter physics~\cite{Ortiz2019}. This material exhibits a dazzling array of exotic phenomena, including superconductivity~\cite{Li2023, Zhong2023, Deng2024, Ge2024, Mu2021}, charge density wave (CDW)~\cite{Yu2021, Guo2022, Zheng2022}, anomalous Hall effect (AHE)~\cite{Yu2021}, pairing density waves (PDW)~\cite{Chen2021}, electronic nematic order~\cite{Tazai2023, Xu2022, Nie2022}, broken time reversal symmetry~\cite{Li2022, Jin2022, Ge2025}. Although, magnetic order is absent and electron correlation is relatively weak in AV$_3$Sb$_5$~\cite{Hu2022, Hu2022a}, it has spurred to search correlated kagome superconduting systems in simialr kagome materials. 

Recently discovered Cr-based kagome metal \ccs~introduces magnetism and electron correlations in the 135 kagome family by substituting Cr atoms for V atoms, enriching their physical properties~\cite{Liu2024}. The \ccs~bulk single crystals undergo a magnetic phase transition around 55 K revealed by magnetic susceptibility and nuclear magnetic resonance, accompanying with a structural modulation identified via single-crystal X ray-diffraction (XRD) results~\cite{Liu2024}. A superconducting transition was observed under high pressure, with the suggested density-wave-like orders are suppressed. Theoretical band calculations predict the flat bands to be closer to the Fermi level (\ef) compared with \cvs, which can serve as origin of strong electron correlation and intertwined orders~\cite{Xie2025, Wu2025}. The coexistence of charge/spin density wave, superconductivity and magnetism in \ccs~has placed the system at the forefront of research in the kagome materials. In particular, the flat bands and their energy positions have been a primary focus in \ccs, as they could be closely related to electron correlation, magnetism, quantum criticality and superconductivity. Previous angle-resolved photoemission spectroscopy (ARPES) experiments with relatively blurred data reported the flat bands lie in the vicinity of or below \ef, implying their role in determining the physical properties of \ccs~\cite{Li2025, Wang2025, Peng2024}. However, this is hard to reconcile with theoretical calculations, which predict the flat band to be above \ef~with and without considering electron correlation effects~\cite{Li2025, Xie2025}. Besides, the microscopic origin of superconductivity and its relationship with the charge order remains elusive as atomic-scale investigations are lacking. Therefore, a comprehensive study of the electronic properties of \ccs~in both the real and momentum space would highly valuable in addressing these open questions.

In this work, we combine scanning tunneling microscopy (STM), ARPES, and density functional theory (DFT) calculations to elucidate the real space charge order and electronic properties of the kagome metal \ccs. Strikingly, we identify new stripe orders on both the Cs- and Sb- terminated surfaces, which are distinct from the bulk stripe order. The kagome flat bands are found to be about 330 meV above \ef. ARPES spectra reveal that the electron correlations in \ccs~originate from Coulomb interactions and Hund's coupling, rather than the kagome flat band. We also discover a distinct electron-boson coupling at 100 meV, different from that in \cvs. Temperature-dependent measurements of the Sb $p$-orbital band show negligible evolution across the phase transition, indicating weak coupling to the order. Our results reveal new surface stripe orders and clarify the nature of many-body interactions in \ccs, offering crucial insights into its flat band charge order and superconductivity.

\section{\label{sec:level1}Results}

\begin{figure}
\includegraphics[width=0.9\textwidth]{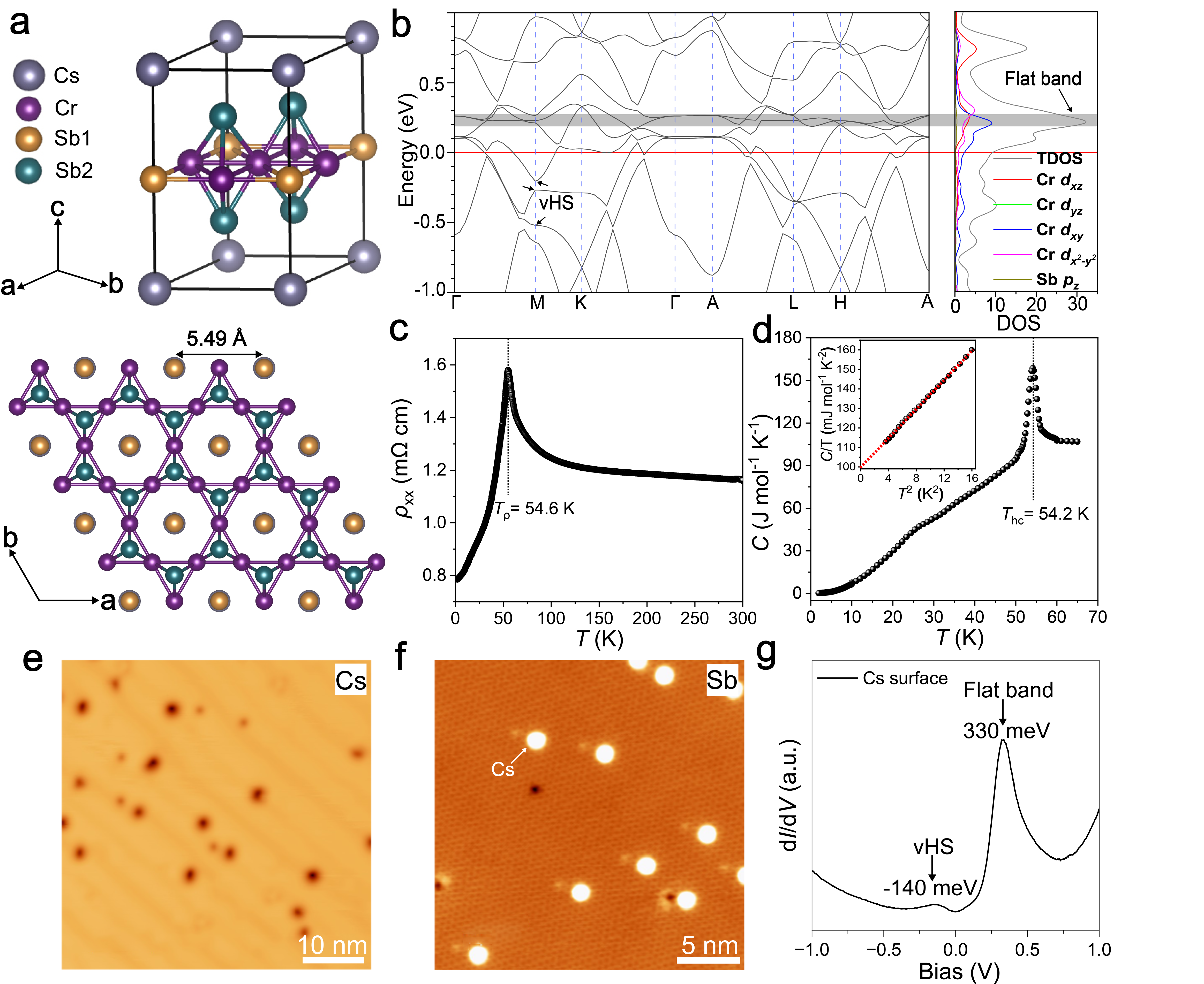}
\caption{\label{fig:Fig1} {\bf Crystal structure, electronic band structure and temperature-driven phase transition of \ccs~single crystals.} \textbf a, Hexagonal crystal structure of \ccs~from both the axonometric and top view. \textbf b, Electronic band structure and the corresponding electronic density of states (DOS) of \ccs~from the first-principles calculations. The van Hove singularities (vHS) are indicated by the black arrows and the flat bands are highlighted by the grey box. \textbf c, Temperature-dependent in-plane resistivity of \ccs~single crystal. The bad metallic behavior at high temperatures is interrupted by a pronounced peak at $T_{\rho}$ = 54.6 K, which relates to an antiferromagnetic transition and structural modulation. \textbf d, Specific heat ($C$) of \ccs, showing a sharp peak at 54.2 K. Inset is a plot and corresponding linear fit of $C/T$ versus $T^2$. \textbf e and \textbf f, STM topographies taken on the Cs- and Sb-terminated surfaces, respectively. Dark dots on the Cs termination indicate Cs vacancies while bright dots on the Sb termination correspond to Cs adatoms. \textbf g, Typical dI/dV spectrum measured on the Cs-terminated surfaces. The peaks at –140 and 330 meV can be attributed to the vHS and flat band, respectively, in alignment with the theoretical calculations. STM setup condition: $V_{s}$ = 2.5 V, $I_{t}$ = 100 pA (e); $V_{s}$ = 1 V, $I_{t}$ = 500 pA (f).
}
\end{figure}

Single crystals of \ccs~were grown utilizing the self-flux method~\cite{Liu2024, Wang2025}. The single crystal displays a lamellar morphology, a silvery metallic luster and a hexagonal shape. We examined the as-synthesized crystals using a powder XRD and energy-dispersive X-ray Spectroscopy (EDS) (Supplementary Fig. 1a-b). These characterization results are consistent with the centrosymmetric structure of \ccs, which is isostructural to AV$_3$Sb$_5$, crystallizing in the space group P6/mmm (No. 191)~\cite{Ortiz2019}. It comprises stacked Cr$_3$Sb$_5$ layers and Cs layers along the \textit{c}-axis (Fig.~\ref{fig:Fig1}a). In the Cr-Sb layers, chromium atoms establish a two-dimensional kagome lattice with Sb1 atoms situated at the heart of the hexagons. The Sb2 atoms establish a two-dimensional honeycomb lattice above and below the kagome layer. The calculated electronic structures and the associated density of states (DOS) are presented in Fig.~\ref{fig:Fig1}b, showing vHS at the M point below the Fermi level and flat bands located above \ef.

We first perform systematic physical property measurements for the bulk single crystals. Figure~\ref{fig:Fig1}c illustrates the temperature dependence of the longitudinal resistivity $\rho_{ab}$ at zero magnetic field. The resistivity exhibits bad metallic behavior at relatively high temperatures and shows a peak at $T_{\rho}$ = 54.6 K followed by an abrupt decline at low temperatures. Corresponding peak anomalies are also evident in the measured specific heat \textit{C(T)} (Fig.~\ref{fig:Fig1}d) and magnetic susceptibility $\chi_{ab}$ (Supplementary Fig. 1c). Analysis of the specific heat data via the formula $C/T = \gamma + \beta T^2$ yields an electron-specific heat coefficient $\gamma_{exp}$ = 100(1) $mJ\cdot K^{-1}\cdot mol^{-1}$. This enhanced value, compared to the results from the AV$_3$Sb$_5$ system~\cite{Liu2024, Ortiz2019}, signaling strong electron correlations in \ccs~(Supporting Fig. 1d-e). These unconventional properties are consistent with previous reports on high-quality \ccs~crystals, which attribute the anomalies to the transition into antiferromagnetic and density-wave orders~\cite{Liu2024}. 

To directly identify the electronic ground state of \ccs, we further investigate the cleaved surfaces using a low-temperature STM. Owing to the weaker bonds between Cs and Sb2 layers, the cleavage planes are typically Cs-terminated or Sb-terminated, as shown in Fig.~\ref{fig:Fig1}e and \ref{fig:Fig1}f. The structural instability is inherent to the Cs-terminated surface, which drives the formation of vacancy defects as observed (Fig.~\ref{fig:Fig1}e). In contrast, the Sb-terminated surface is prone to the adsorption and aggregation of excess Cs adatoms (Fig.~\ref{fig:Fig1}f). The measured height between the adjacent Cs and Sb layers is about 1.46 nm (Supplementary Fig. 2a-b), consistent with the height of the upper Cs and lower Sb layers in adjacent layers. We conducted differential conductance spectroscopy on both the Cs- and Sb-terminated surfaces. The $dI/dV$ spectrum obtained from defect-free areas exhibits a prominent peak at 330 meV and a bulge at −140 meV on the Cs plane (Fig.~\ref{fig:Fig1}g). Aligning with the theoretical results, the bulge at −140 meV is indicative of the M-point vHS, whereas the peak at 330 meV is associated with the flat electronic band, as will be discussed later. Meanwhile, the $dI/dV$ spectrum obtained on the defect-free Sb surface also exhibits peak from the vHS at the M point (Supplementary Fig. 3). 

\begin{figure}
\includegraphics[width=0.95\textwidth]{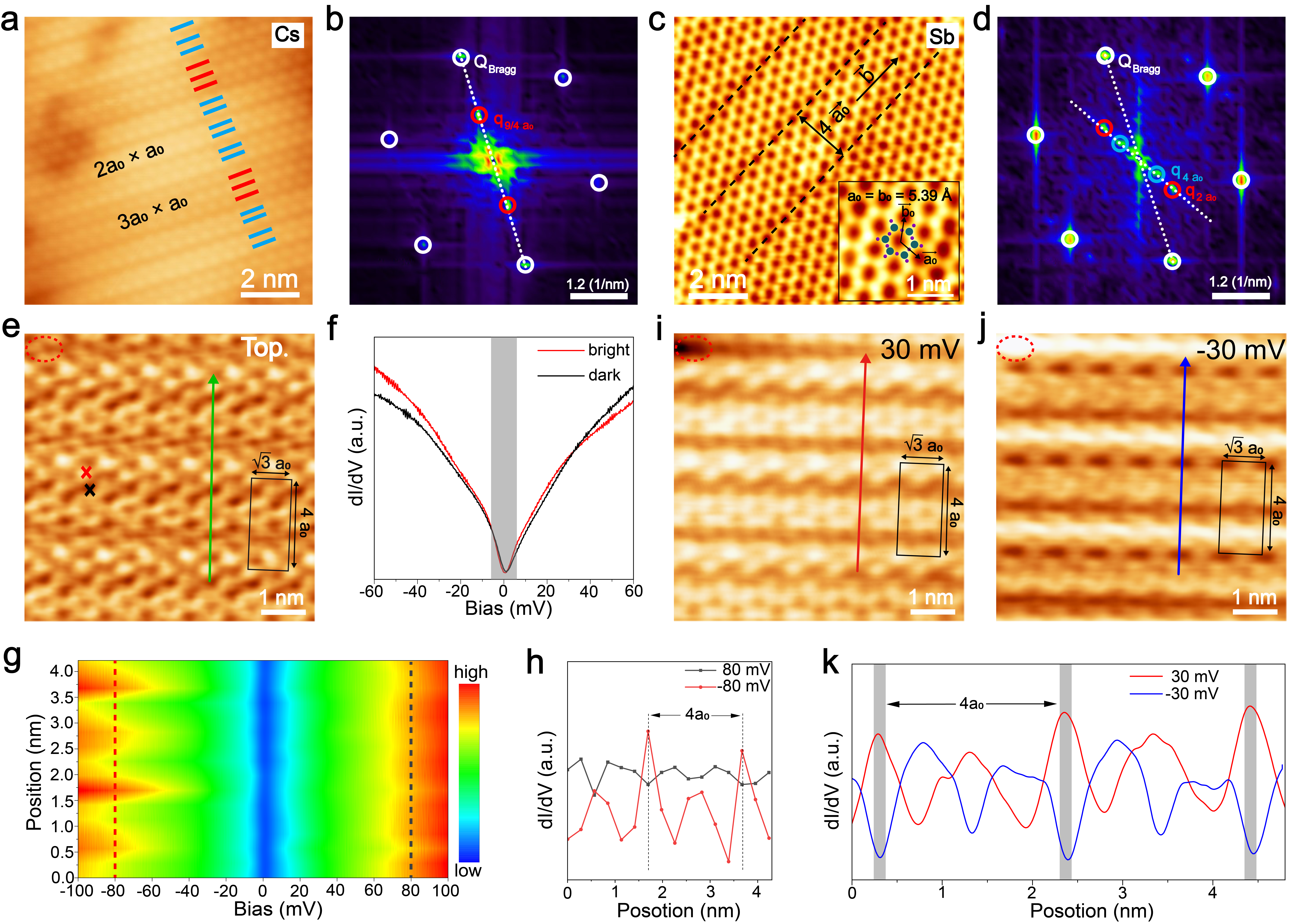}
\caption{\label{fig:Fig2} {\bf Distinct surface stripe orders of \ccs.} \textbf a, High-resolution STM topography taken on the Cs-terminated surface with a mixture of 2$a_0$×$a_0$ and 3$a_0$×$a_0$ stripe superstructures. Every three 2a$_0$ stripes are separated by a 3a$_0$ stripe. \textbf b, Fast Fourier transform (FFT) of \textbf a. Superstructure modulation peak shows at 4/9 $Q_{Bragg}$. \textbf c, High-resolution STM topography taken on the Sb-terminated surface with a 4$a_0$×$\sqrt{3}$ $a_0$ stripe superstructure. The illustration shows the honeycomb structure with a lattice constant of 5.39 \AA. \textbf d, FFT pattern of \textbf c. The $q_{2a_0}$ and $q_{4a_0}$ are rotated by 30° relative to $Q_{Bragg}$. \textbf e, The STM topography of the Sb-terminated surface with a 4$a_0$ stripe superstructure. \textbf f, The dI/dV spectra were acquired from the bright and dark regions at the marked cross in (e). \textbf g, A series of dI/dV spectra along the green arrow in (e) encompasses two 4$a_0$ periods. \textbf h, The spatial evolution of the dI/dV intensity, extracted from the panel (g) at ±80 mV, shows a clear 4$a_0$ periodicity, with the peaks at −80 mV and the troughs at 80 mV aligning in phase. \textbf {i and j,} dI/dV mapping at 30 mV (i) and −30 mV (j), a defect is marked with a red dotted circle. \textbf k, Line profiles along the same region, marked by arrows in (i) and (j), showing the peak-to-dip intensity contrast with a 4$a_0$ period. 4$a_0$ $\times$ $\sqrt{3}a_0$ stripe superstructure is marked as a black rectangle in (e), (i) and (j). STM setup condition: (a) $V_s$ = 130 mV, $I_t$ = 1.0 nA; (c) $V_s$ = −300 mV, $I_t$ = 100 pA; (e) $V_s$ = −30 mV, $I_t$ = 1.0 nA; (i) $V_s$ = 30 mV, $I_t$ = 1.0 nA; (j) $V_s$ = −30 mV, $I_t$ = 1.0 nA.
}
\end{figure}

We obtained high quality atomically resolved STM topographies which correspond to the Cs- and Sb-terminated surfaces, showing hexagonal and honeycomb lattices with measured lattice constant of 5.40 \AA~and 5.39 \AA~(Figs.~\ref{fig:Fig2}a and 2c). On the Cs-terminated surface, a mixed superstructure composed of 2$a_0$  $\times$ $a_0$  and 3$a_0$  $\times$ $a_0$ stripe patterns is observed. Every 3$a_0$ stripe is separated by three 2$a_0$ stripes and the corresponding fast Fourier transform (FFT) map (Fig.~\ref{fig:Fig2}b) exhibits peaks at $q_{9/4a_0}$ between $q_{2a_0}$ and $q_{3a_0}$ along the $Q_{Bragg}$ lattice direction, further confirming these periodic modulations. We notice that the Cs atoms are compactly arranged without any vacancies on the surface exhibiting mixed stripe orders. In addition, the stripes with distinct orientations show smooth transitions at their intersection regions (Supplementary Fig. 4). A suppression of the DOS near \ef~is also observed in the local $dI/dV$ spectra (Supplementary Fig. 4). These observations suggest that the stripe phase on the Cs-terminated surface arises from a charge order rather than a simple surface reconstruction. Interestingly, the stripe order is distinct from the stripe phase observed in the bulk of \ccs~(4$a_0\times a_0$). 
On the Sb-terminated surface (Fig.~\ref{fig:Fig2}c), a 4$a_0\times\sqrt{3}a_0$ stripe structure emerges, where the vectors along the $\sqrt{3}a_0$ direction are rotated by 30° relative to the crystal lattice. This also differs from the 4$a_0\times a_0$ structure observed in bulk \ccs~via single-crystal XRD and optical measurements~\cite{Liu2024,Meta2024}. Under the specific tip condition, the sample shows a densely packed atomic-resolved pattern and the 4$a_0\times\sqrt{3}a_0$ periodic stripe structure becomes more pronounced (Supplementary Fig. 5a). This stripe order is further confirmed by the FFT analysis (Fig.~\ref{fig:Fig2}d and Supplementary Fig. 5b) of the topography, which exhibits vector peaks corresponding to $q_{2a_0}$ and $q_{4a_0}$. Additionally, we observed 4$a_0$ $\times$ $\sqrt{3}a_0$ stripe patterns with different orientations in a large area (Supplementary Fig. 6), indicating their prevalent nature on the Sb-terminated surface.

We found that the 4$a_0$ $\times$ $\sqrt{3}a_0$ stripe phase exhibits bias-dependent characteristics (Supplementary Fig. 7). To further elucidate the origin of the stripe phase, we measured $dI/dV$ spectra from both the bright and dark regions of the pattern, as shown in Figs.~\ref{fig:Fig2}e and \ref{fig:Fig2}f. Both regions exhibit an energy gap of approximately 10 meV (Fig.~\ref{fig:Fig2}f and Supplementary Fig. 8). Notably, the $dI/dV$ intensity is higher in the bright region under higher negative bias, whereas the dark region prevails under higher positive bias. Furthermore, a series of $dI/dV$ spectra were measured along a direction perpendicular to the stripe order (green arrow in Fig.~\ref{fig:Fig2}g). The spatial modulation period is approximately 2.1 nm, which aligns with the unidirectional $4a_0$ charge order. Simultaneously, we also observe that the $4a_0$ charge order exhibits a peak-to-valley feature under opposite bias voltages (Fig.~\ref{fig:Fig2}h). By switching the bias from positive to negative value but keeping other parameters same, we observe a clear contrast inversion in the $dI/dV$ mapping of the same region, as demonstrated in Figs.~\ref{fig:Fig2}i and \ref{fig:Fig2}j. The $dI/dV$ intensity profile along the same position exhibits a well-defined $4a_0$ periodic modulation with a peak-to-valley feature at ±30 mV (Fig.~\ref{fig:Fig2}k). These results strongly suggest it to be a stripe CDW order.

\begin{figure}
\includegraphics[width=0.98\textwidth]{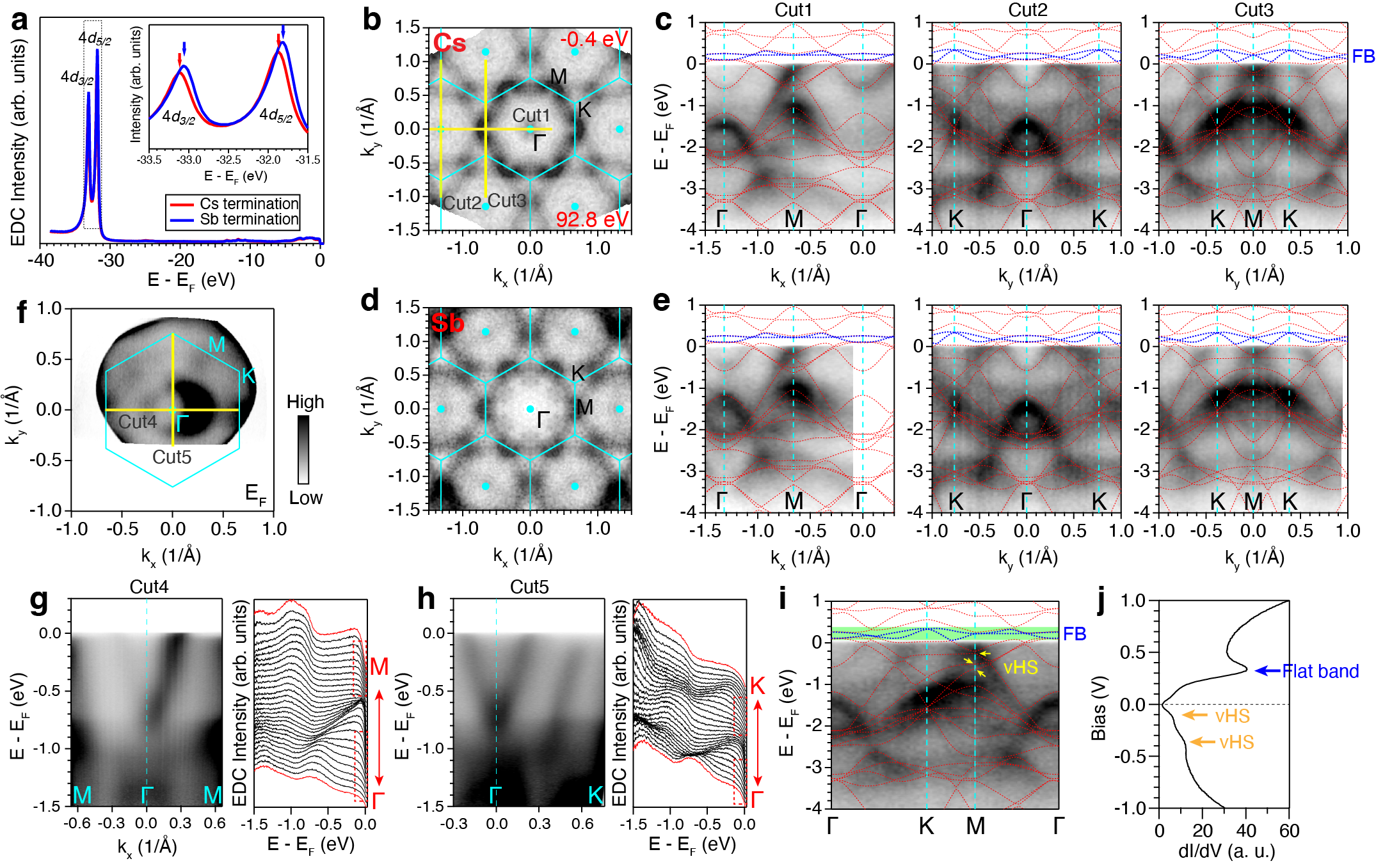}
\caption{\label{fig:Fig3} {\bf Momentum-resolved electronic structure and the energy position of kagome flat bands of \ccs.} \textbf a, Photoemission spectra of \ccs~measured with 92.8 eV photons, showing the Sb 4$d$ core level, which indicates different surface terminations. \textbf b, Constant energy contour mapping at -0.4 eV on the Cs-terminated surface obtained using circularly polarized light. \textbf c, Spectral images along different momentum paths indicated by the yellow lines in \textbf b. The overlaid dashed lines represent the corresponding calculated bands and the blue dashed lines highlight the kagome flat bands. \textbf {d and e}, Same as \textbf{b,c} but for the Sb-terminated surface. \textbf f, Fermi surface mapping measured with a 7 eV laser light source. A sample bias of –50 eV was applied to achieve a large momentum detection range. The light polarization was linear horizontal. \textbf g, $E-k$ spectral image and the corresponding energy distribution curve (EDC) stacks along the horizontal cut (cut4) across $\Gamma$, as indicated in \textbf f. The red dashed boxes highlight the clean Fermi cutoff without any flat-band spectral feature. \textbf h, Same as \textbf g but along the vertical cut (cut5) indicated in \textbf f. \textbf i, ARPES spectral image overlaid with DFT vHS. \textbf j, dI/dV spectra obtained by combining data from the Cs- and Sb-terminated surfaces. Flat bands and vHS are labeled with the blue and yellow arrows, respectively.
}
\end{figure}

Having identified distinct surface stripe orders in \ccs, we now turn to explore their underlying electronic origins. Since cleaving the crystal produces two different surface terminations, and our STM/STS studies revealed distinct surface stripe patterns, we employed nano-ARPES with a submicron spot size ($<$ 1 $\mu$m) to probe the corresponding electronic structures. The two surface terminations can be distinguished by the Sb 4$d$ core-level shifts, which reflect their different crystal environments (Fig.~\ref{fig:Fig3}a)~\cite{Kato2023}. The constant energy contour at -0.4 eV on the Cs-terminated surface exhibits a characteristic Star-of-David Fermi surface at the van Hove filling of a kagome lattice~\cite{Kiesel2013a, Hu2022a}, along with a small electron pocket around the Brillouin zone center which mainly consists of Sb 5$p$ orbitals (Fig.~\ref{fig:Fig3}b). The ARPES $E-k$ spectral images along different high-symmetry directions show good agreement with the first-principles calculations without including the surface reconstruction effects (Fig.~\ref{fig:Fig3}c). We notice a slight energy discrepancy between the calculated and measured positions of the hole-like band at around -1.5 eV near $\Gamma$. However, this difference can not be accounted for by simple electron-correlation–induced band renormalization, as such an effect would shift the calculated bands toward \ef, thus further enlarge the mismatch. Nevertheless, the first-principles calculations capture the overall electronic structure of \ccs~well. An examination on the Sb-terminated surface by ARPES reveals essentially the same electronic structure as that of the Cs-terminated surface (Figs.~\ref{fig:Fig3}d and 3e). Considering that the ARPES spectra of \ccs~are not as sharp as those of \cvs~\cite{Hu2022a}, any subtle differences in the electronic structure may be smeared out by the broader spectral features in \ccs, rendering them indistinguishable. Since the measured electronic structures on the two different surface terminations of \ccs~are indistinguishable, we will not differentiate between them afterwards. 

Previously, flat bands were reported at or below \ef~in \ccs~with the extra spectral peak observed around \ef~away from the dispersive bands~\cite{Li2025, Wang2025, Peng2024}. In contrast, our ARPES $E-k$ spectral images show no such features. Instead, they exhibit a clean cutoff in the background at \ef~away from the dispersive bands (Figs.~\ref{fig:Fig3}c and 3e). The measurements were conducted with circularly polarized light, which should not significantly suppress possible flat band signals. To further examine the spectral weight near \ef, we carried out high-resolution laser-ARPES measurements on \ccs~(Figs.~\ref{fig:Fig3}f-h). The $E-k$ spectral images along both $\Gamma -$ M and $\Gamma -$ K clearly resolve the electron band centered at the $\Gamma$ point, showing sharp contrast against the background (Figs.~\ref{fig:Fig3}g and 3h). However, the corresponding energy distribution curves display no additional peaks at \ef~outside the dispersive band region (red dashed boxes in Figs.~\ref{fig:Fig3}g and 3h). These observations unambiguously rule out the presence of flat bands in the vicinity of \ef.

The question is how high the flat bands lie above \ef. Although the flat bands above \ef~cannot be directly observed in ARPES due to the Fermi cutoff, their positions can be inferred from the band calculations, which locate about 200 meV above the Fermi level (Fig.~\ref{fig:Fig1}b). Given the overall good agreement between the calculations and our ARPES data, it is reasonable to assign the flat-band position accordingly (Fig.~\ref{fig:Fig1}b and Figs.~\ref{fig:Fig3}c and 3e). Consistently, our STS spectra show a pronounced peak at approximately 330 meV above \ef~( Fig.~\ref{fig:Fig3}j), corresponding to the flat band energy. The combined evidence from ARPES, STM, and DFT calculations thus reconciles the previous discrepancies between theoretical predictions and experimental observations regarding the flat-band position in \ccs~\cite{Xie2025, Li2025, Wang2025, Peng2024}.

\begin{figure}
\includegraphics[width=0.95\textwidth]{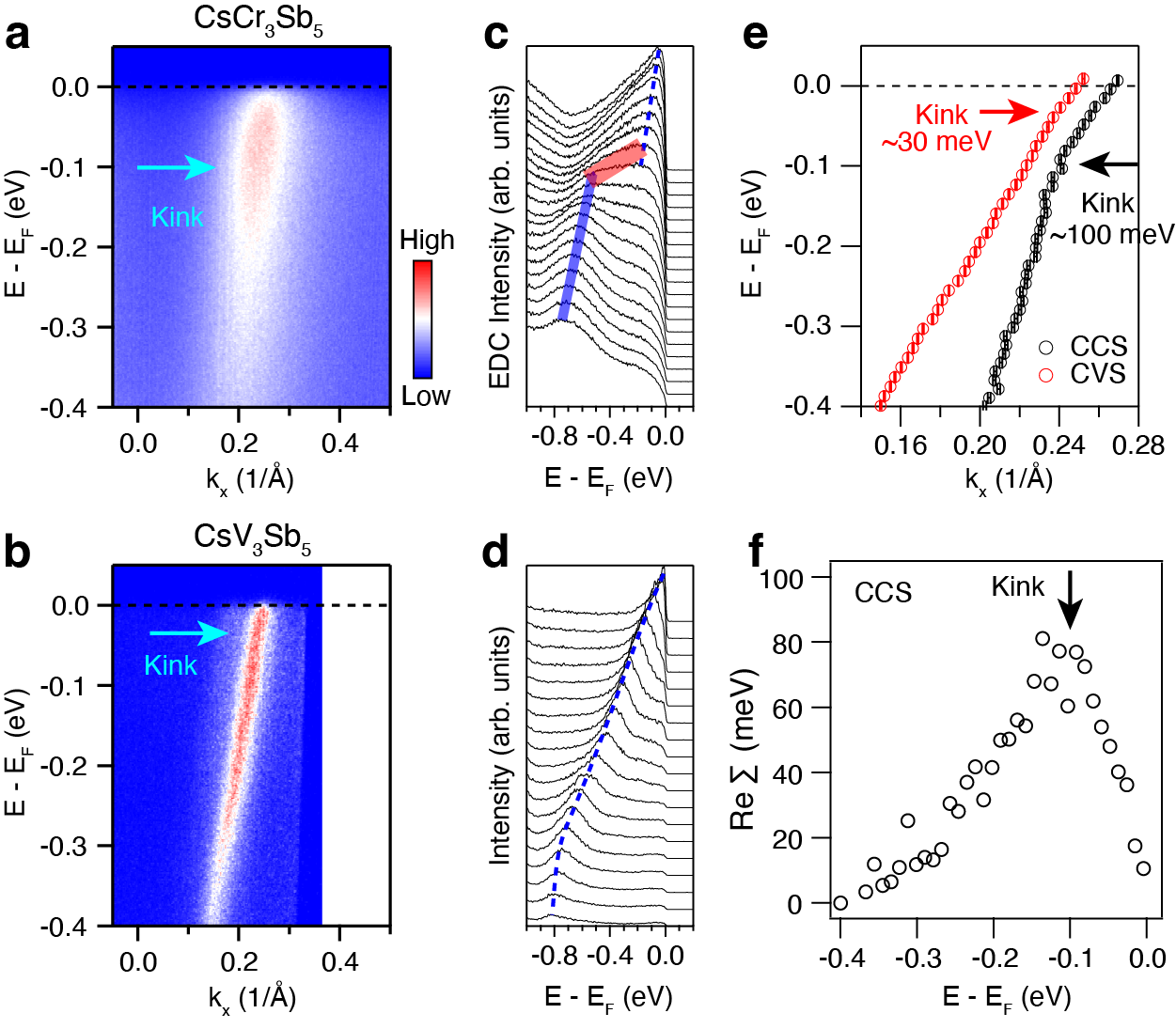}
\caption{\label{fig:Fig4} {\bf Electron correlation and distinct electron-boson coupling in \ccs.} \textbf a, $E-k$ spectral image of \ccs~along the cut indicated in Fig.~\ref{fig:Fig2}f (cut4). \textbf b, Same as \textbf a but for \cvs. \textbf {c and d}, Energy distribution curve stacks corresponding to the spectral images in \textbf a and \textbf b, respectively. The blue dashed lines indicate the quasiparticle peaks. The solid blue line denotes incoherent spectra at high binding energies. The red shaded line indicates the waterfall-like spectral feature. \textbf e, Fitted band dispersions of \ccs~(black circles) and \cvs~(red circles) from \textbf a and \textbf b. Kink positions are denoted by the arrows. \textbf f, Real part of the self-energy of \ccs~extracted from \textbf e, assuming a linear bare-band dispersion.
}
\end{figure}

Since the flat bands lie approximately 330 meV above \ef, they are unlikely to drive the strong electron correlation effects in \ccs. To further investigate the origin of electron correlations, we compared the electronic bands of \ccs~and \cvs~measured using a 7 eV laser (Figs.~\ref{fig:Fig4}a and 4b). The $E-k$ spectral image of \cvs~appears much shaper than that of \ccs. Interestingly, the spectral image of \ccs~resembles that of cuprates and iron-based superconductors~\cite{Zhang2008a, Chang2024}. To further visualize this, we plot their corresponding EDC stacks in Figs.~\ref{fig:Fig4}c,d. The EDC stacks of \ccs~exhibit sharp peaks near the Fermi level corresponding to quasiparticle peaks, as well as incoherent peaks around -0.7~eV. A waterfall-like feature connects the quasiparticle peaks and the incoherent peaks (red shaded line in Figs.~\ref{fig:Fig4}c), a hallmark of correlated systems with strong Coulomb interactions and Hund's coupling~\cite{Krsnik2025}. In contrast, the EDCs stacks of \cvs~show quasiparticle peaks extending continuously from the Fermi level to a high binding energy of -0.8~eV, characteristic of a weakly correlated system. These results suggest that \ccs~is a strongly correlated system, with correlations arising from Coulomb interactions and Hund's coupling. Moreover, fitting the corresponding band dispersion of \ccs~in Fig.~\ref{fig:Fig4}a reveals a kink at approximately -100 meV (Figs.~\ref{fig:Fig4}e and 4f), distinct from the $\sim$30 meV kink observed in \cvs. The $\sim$30 meV kink in \cvs~was attributed to a phonon mode~\cite{Zhong2023a}. Given the potential magnetic nature of \ccs, the 100 meV mode is unlikely to be phononic and is more likely associated with spin excitations.
 
\begin{figure}
\includegraphics[width=0.95\textwidth]{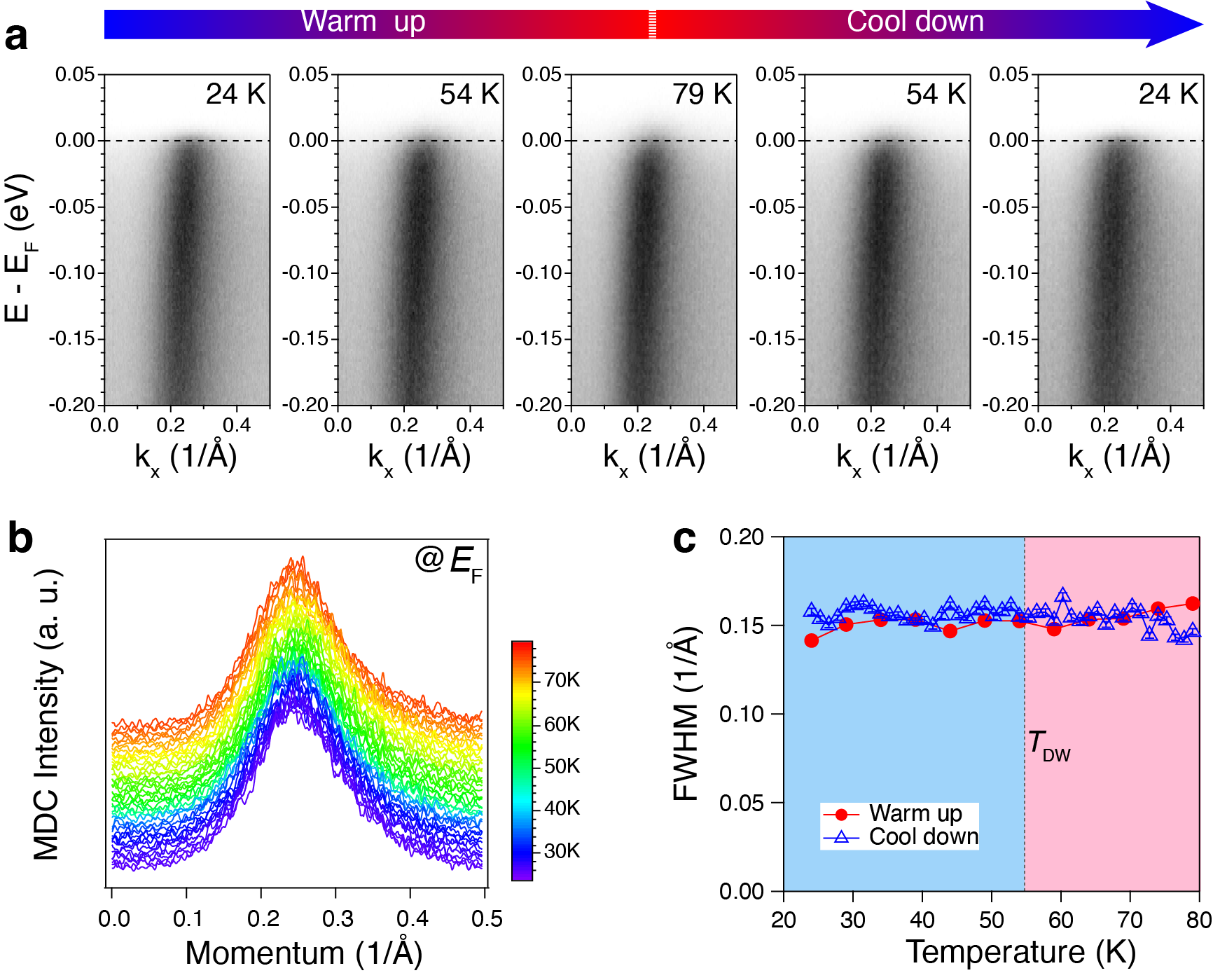}
\caption{\label{fig:Fig5} {\bf Temperature evolution of the electronic band originating from the Sb $p$ orbital.} \textbf a, ARPES spectral images of the band along the cut indicated in Fig.~\ref{fig:Fig2}f (cut4) measured at different temperatures. Warm-up and cool-down cycles were performed to rule out extrinsic effects introduced by sample degradation. \textbf b, Corresponding momentum distribution curves at the Fermi level of the $E-k$ spectra shown in \textbf a for the cool-down process. \textbf c, Fitted peak width of \textbf b, showing no significant change across the transition temperature for either the warm-up or cool-down processes.
}
\end{figure}

Our laser-ARPES data on \ccs~reveal the clear electron pocket around $\Gamma$, prompting us to investigate its response to the sample's phase transition (Fig.~\ref{fig:Fig5}). This electron band primarily consists of Sb $p$ orbitals~\cite{Wang2025}. We performed detailed temperature-dependent measurements with both warming and cooling cycles (Fig.~\ref{fig:Fig5}a and Supplementary Fig. 9). The $E-k$ spectral images show no significant changes with temperature, as further confirmed by the corresponding momentum distribution curves (MDCs) at \ef~in Fig.~\ref{fig:Fig5}b. The MDCs at different temperatures exhibit similar shapes without abrupt changes in peak height or width, in contrast to a previous report~\cite{Li2025}. To quantify this, a single-peak fit of the MDCs were performed. The corresponding fitting result shows no change in peak width across the phase transition for both warming and cooling cycles (Fig.~\ref{fig:Fig5}c). We note that the peak width at high temperature (79 K) is slightly smaller than that at low temperature (24 K) in the cooling cycle, likely due to minor surface degradation over time. The temperature-independent behavior of the Sb $p$ orbital band suggests that it couples weakly to the bulk density wave order. Thus the bulk density wave order likely originates from the Cr $d$ orbitals.

\section{Discussion and Conclusion}
In \cvs, the surfaces develop additional charge orders, which have been attributed to electronic rotational symmetry breaking and pair-density-wave states~\cite{Zhao2021b, Chen2021}. These supermodulations are superimposed on the bulk 2$\times$2 charge order. In contrast, both the Cs- and Sb- terminated surfaces of \ccs~do not inherit its bulk stripe order, but instead develop entirely new surface stripe orders. Remarkably, to the best of our knowledge, such surface stripe orders have never been reported in any kagome material before. Theoretical calculations on the relative energies of different surface configurations would be valuable for understanding the formation mechanism of these stripe patterns. In particular, the $4a_0\times\sqrt{3}a_0$ stripe order effectively enlarges the surface unit cell, which could potentially stabilize altermagnetic states on the surface of \ccs~through appropriately arranged magnetic blocks~\cite{Xu2025}. With these distinct surface stripe orders identified, a key question arises as to whether superconductivity in \ccs~originates from the bulk or the surface. In this regard, investigating how the surface stripe orders evolve under hydrostatic pressure would be crucial for uncovering their relationship with superconductivity.

Since the electronic structures measured on the two different terminations show no discernible differences, we consider them to primarily reflect the bulk electronic properties. Our results reveal that the electron correlations in \ccs~are of a more “traditional” origin, rather than being driven by the kagome flat bands, with electronic spectra resembling those of cuprates and iron-based superconductors~\cite{Zhang2008a, Chang2024}. Moreover, no obvious Fermi surface nesting condition is observed. In this context, we suggest that the stripe orders in \ccs, both in the bulk and on the surface, are closely related to electron correlations. In addition, the observed electron-boson coupling to a $\sim$100 meV mode may also play an important role. We note that the correlated spectra are revealed here by the central electron pocket with a dominant Sb $p_z$ orbital character, indicating non-negligible $p-d$ orbital mixing~\cite{Han2023}.

Despite the flat band being located about 330 meV above \ef, a theoretical study has predicted strong magnetic fluctuations arising from this flat band in \ccs~\cite{Wu2025}. However, our temperature-dependent measurements of the Sb $p_z$ orbital band show no discernible change in the scattering rate across the transition temperature, ruling out enhanced electron scattering from magnetic fluctuations. The abrupt increase in resistivity at the transition temperature could instead result from the suppression of electronic states near the Fermi level, as revealed by our STS measurements. Nevertheless, the Sommerfeld coefficient remains nearly six times larger than that predicted by DFT calculations~\cite{Liu2024}, highlighting the strong correlation effects in \ccs.

To summarize, we have discovered new surface stripe orders on different terminations of \ccs. These stripe orders are distinct from the bulk stripe order and have not been reported previously in any kagome lattice material. We also find that the electron correlation effects in \ccs~are not driven by the kagome flat bands but rather originate from Coulomb interactions and Hund’s coupling, laying the electronic foundation for understanding its exotic phases. In addition, we identify a new coupling mode at approximately 100 meV, accompanied by no observable enhancement in the scattering of Sb $p$ electrons due to magnetic fluctuations across the phase transition. Our findings open new avenues for understanding the relationship and interplay between surface and bulk electronic states in generating exotic orders in strongly correlated kagome systems.

During the preparation of this manuscript, two independent studies appeared on arXiv, reporting the 4$a_0\times\sqrt{3}a_0$ stripe order on the Sb-terminated surface of \ccs~\cite{Huang2025a, Cheng2025}.

\section{Methods}
\subsection{\label{sec:level2} Sample growth and characterization}
Single crystals of \ccs~were synthesized using the flux method. Stoichiometric quantities of Cs liquid (99.98$\%$, Alfa), Cr powder (99.95$\%$, Alfa), and Sb ingot (99.99$\%$, Trillion Metals Co) were amalgamated in a molar ratio of 4:1:9, placed into an alumina crucible, and hermetically sealed within a Nb/Ta tube under an argon environment at 0.2 atm utilizing arc welding. To avert oxidation, the Nb/Ta tube was additionally encased within an evacuated quartz ampoule. The assembly was incrementally heated to 1173 K and sustained at this temperature for 24 hours to guarantee thorough homogeneity of the melt. The material was subsequently chilled to 873 K at a regulated rate of 2 $^\circ$C/h, and then further naturally cooled to room temperature within the furnace. The resultant product was submerged in deionized water, producing van der Waals hexagonal platelets with typical lateral dimensions of around $1 \times 1$ mm$^2$.

Crystal structure characterization was conducted at ambient temperature via XRD utilizing Cu K$\alpha$ ($\lambda$ = 1.5406 \AA) radiation on a PANalytical Empyrean Series 2 diffractometer. Scanning electron microscopy (SEM) and EDS studies were conducted utilizing a ZEISS EVO MA 10 and a BRUKER XFlash. Measurements of electrical resistivity and specific heat were conducted with a Physical Property Measurement System (PPMS-9, Quantum Design). The magnetic characteristics were assessed using a Magnetic Property Measurement System (MPMS, Quantum Design).

\subsection{\label{sec:level2} ARPES measurements}
Synchrotron-based ARPES measurements were carried out at the BL07U endstation of the Shanghai Synchrotron Radiation Facility (SSRF), using a Scienta DA30L electron analyzer. The base pressure during measurements was lower than $1 \times 10^{-10}$~mbar. All samples were cleaved $in~situ$ at 300~K in a preparation chamber with a base pressure of $1 \times 10^{-10}$~mbar, and subsequently transferred to the measurement chamber. The measurement temperature was maintained at 22~K unless otherwise specified in the main text. A focused beam spot better than $1 \times 1$ $\mu$m$^2$ was achieved using a Fresnel zone plate. The measurements were performed with a photon energy of 92.8 eV and circularly polarization light. The overall energy resolution was $\sim$20 meV and the angular resolution was $\sim$0.3$^\circ$.

Laser-ARPES measurements were performed at the Institute of Physics, Chinese Academy of Sciences, using a 7~eV laser light source combined with a Scienta DA30L electron Analyzer. All samples were cleaved $in~situ$ at 24~K. The base pressure was lower than $5 \times 10^{-11}$~mbar during measurements. The laser beam spot was less than $10 \times 10$ $\mu$m$^2$ using a focusing lens. The overall energy resolution was $\sim$3 meV unless otherwise specified in the main text, and the angular resolution was $\sim$0.3$^\circ$.

\subsection{\label{sec:level2} Scanning tunneling microscopy/spectroscopy}
Single crystals of \ccs~were cleaved $in~situ$ under ultrahigh vacuum (UHV) at a temperature of 10 K and subsequently transferred directly into STM chamber for investigations. Experiments were performed in an UHV ($1 \times 10^{-10}$ mbar) low-temperature STM system (Unisoku USM-1300). Typical STM/STS results were measured at 5.2 K. All STM measurements were performed with constant current mode using a PtIr tip. The dI/dV spectra were measured by using the lock-in technique with a reference signal at 911 Hz. The modulation amplitudes were set as 50 mV for the large bias range and 3 $\sim$ 5 mV for the small bias range. The STM images in this work were processed using the WSxM software~\cite{Horcas2007}.

\subsection{\label{sec:level2} DFT calculations}
The structure of \ccs~was optimized to preserve periodic boundary conditions while minimizing spurious interactions from periodic images. Density functional theory (DFT) calculations within the generalized gradient approximation (GGA-PBE)~\cite{Perdew1996} were performed using the Vienna Ab Initio Simulation Package (VASP)~\cite{Kresse1996}, employing a 600 eV plane-wave cutoff and 11 $\times$ 11 $\times$ 11 Monkhorst-Pack $k$-mesh to ensure energy convergence ($<$ 1 meV/atom) and residual forces ($<$ 1 meV/\AA).

\section{acknowledgments}
This work is financially supported by the Guangdong Provincial Quantum Science Strategic Initiative (Grant No. GDZX2401009), National Natural Science Foundation of China (Grant No. 21BAA01133, 12374052, 12474182, 92165204),  Guangdong Basic and Applied Basic Research Foundation (Grant No. 2023A1515010487), Guangzhou Basic and Applied Basic Research Foundation (Grant No. 2025A04J5405), Research Center for Magnetoelectric Physics of Guangdong Province (2024B0303390001), Guangdong Provincial Key Laboratory of Magnetoelectric Physics and Devices (Grant No. 2022B1212010008), the China Postdoctoral Science Foundation (Grant No. 2024M763734) and the Instrumental Analysis \& Research Center, Sun Yat-sen University.

\section{Data Availability}
All data needed to evaluate the conclusions are present in the paper and supplementary materials. Additional data are available from the corresponding authors on reasonable request.

\section{Competing interests}
The authors declare no competing interests. 


%\nocite{*}

%\bibliographystyle{naturemag}
%\bibliography{CCS}

\begin{thebibliography}{10}
\expandafter\ifx\csname url\endcsname\relax
  \def\url#1{\texttt{#1}}\fi
\expandafter\ifx\csname urlprefix\endcsname\relax\def\urlprefix{URL }\fi
\providecommand{\bibinfo}[2]{#2}
\providecommand{\eprint}[2][]{\url{#2}}

\bibitem{Yin2022}
\bibinfo{author}{Yin, J.~X.}, \bibinfo{author}{Lian, B.} \&
  \bibinfo{author}{Hasan, M.~Z.}
\newblock \bibinfo{title}{{Topological kagome magnets and superconductors}}.
\newblock \emph{\bibinfo{journal}{Nature}} \textbf{\bibinfo{volume}{612}},
  \bibinfo{pages}{647--657} (\bibinfo{year}{2022}).

\bibitem{Wang2023}
\bibinfo{author}{Wang, Y.}, \bibinfo{author}{Wu, H.},
  \bibinfo{author}{McCandless, G.~T.}, \bibinfo{author}{Chan, J.~Y.} \&
  \bibinfo{author}{Ali, M.~N.}
\newblock \bibinfo{title}{{Quantum states and intertwining phases in kagome
  materials}}.
\newblock \emph{\bibinfo{journal}{Nature Reviews Physics}}
  \textbf{\bibinfo{volume}{5}}, \bibinfo{pages}{635--658}
  (\bibinfo{year}{2023}).

\bibitem{Neupert2022}
\bibinfo{author}{Neupert, T.}, \bibinfo{author}{Denner, M.~M.},
  \bibinfo{author}{Yin, J.-X.}, \bibinfo{author}{Thomale, R.} \&
  \bibinfo{author}{Hasan, M.~Z.}
\newblock \bibinfo{title}{{Charge order and superconductivity in kagome
  materials}}.
\newblock \emph{\bibinfo{journal}{Nature Physics}}
  \textbf{\bibinfo{volume}{18}}, \bibinfo{pages}{137--143}
  (\bibinfo{year}{2022}).

\bibitem{Wilson2024}
\bibinfo{author}{Wilson, S.~D.} \& \bibinfo{author}{Ortiz, B.~R.}
\newblock \bibinfo{title}{{AV$_3$Sb$_5$ kagome superconductors}}.
\newblock \emph{\bibinfo{journal}{Nature Reviews Materials}}
  \textbf{\bibinfo{volume}{9}}, \bibinfo{pages}{420--432}
  (\bibinfo{year}{2024}).

\bibitem{Jiang2023}
\bibinfo{author}{Jiang, K.} \emph{et~al.}
\newblock \bibinfo{title}{{Kagome superconductors AV$_3$Sb$_5$(A = K, Rb,
  Cs)}}.
\newblock \emph{\bibinfo{journal}{National Science Review}}
  \textbf{\bibinfo{volume}{10}}, \bibinfo{pages}{nwac199}
  (\bibinfo{year}{2023}).

\bibitem{Zhang2025}
\bibinfo{author}{Zhang, Z.} \emph{et~al.}
\newblock \bibinfo{title}{{2D Kagome Materials: Theoretical Insights,
  Experimental Realizations, and Electronic Structures}}.
\newblock \emph{\bibinfo{journal}{Advanced Functional Materials}}
  \textbf{\bibinfo{volume}{35}}, \bibinfo{pages}{2416508}
  (\bibinfo{year}{2025}).

\bibitem{Ortiz2019}
\bibinfo{author}{Ortiz, B.~R.} \emph{et~al.}
\newblock \bibinfo{title}{{New kagome prototype materials: discovery of
  KV$_3$Sb$_5$, RbV$_3$Sb$_5$, and CsV$_3$Sb$_5$}}.
\newblock \emph{\bibinfo{journal}{Physical Review Materials}}
  \textbf{\bibinfo{volume}{3}}, \bibinfo{pages}{094407} (\bibinfo{year}{2019}).

\bibitem{Li2023}
\bibinfo{author}{Li, H.} \emph{et~al.}
\newblock \bibinfo{title}{{Small Fermi Pockets Intertwined with Charge Stripes
  and Pair Density Wave Order in a Kagome Superconductor}}.
\newblock \emph{\bibinfo{journal}{Physical Review X}}
  \textbf{\bibinfo{volume}{13}}, \bibinfo{pages}{031030}
  (\bibinfo{year}{2023}).

\bibitem{Zhong2023}
\bibinfo{author}{Zhong, Y.} \emph{et~al.}
\newblock \bibinfo{title}{{Nodeless electron pairing in CsV$_3$Sb$_5$-derived
  kagome superconductors}}.
\newblock \emph{\bibinfo{journal}{Nature}} \textbf{\bibinfo{volume}{617}},
  \bibinfo{pages}{488--492} (\bibinfo{year}{2023}).

\bibitem{Deng2024}
\bibinfo{author}{Deng, H.} \emph{et~al.}
\newblock \bibinfo{title}{{Chiral kagome superconductivity modulations with
  residual Fermi arcs}}.
\newblock \emph{\bibinfo{journal}{Nature}} \textbf{\bibinfo{volume}{632}},
  \bibinfo{pages}{775--781} (\bibinfo{year}{2024}).

\bibitem{Ge2024}
\bibinfo{author}{Ge, J.} \emph{et~al.}
\newblock \bibinfo{title}{{Charge-4e and Charge-6e Flux Quantization and Higher Charge Superconductivity in Kagome Superconductor Ring Devices}}.
\newblock \emph{\bibinfo{journal}{Physical Review X}}
  \textbf{\bibinfo{volume}{14}}, \bibinfo{pages}{021025}
  (\bibinfo{year}{2024}).

\bibitem{Mu2021}
\bibinfo{author}{Mu, C.} \emph{et~al.}
\newblock \bibinfo{title}{{S-Wave Superconductivity in Kagome Metal
  CsV$_3$Sb$_5$ Revealed by $^{121/123}$Sb NQR and $^{51}$V NMR Measurements}}.
\newblock \emph{\bibinfo{journal}{Chinese Physics Letters}}
  \textbf{\bibinfo{volume}{38}}, \bibinfo{pages}{077402}
  (\bibinfo{year}{2021}).

\bibitem{Yu2021}
\bibinfo{author}{Yu, F.~H.} \emph{et~al.}
\newblock \bibinfo{title}{{Concurrence of anomalous Hall effect and charge
  density wave in a superconducting topological kagome metal}}.
\newblock \emph{\bibinfo{journal}{Physical Review B}}
  \textbf{\bibinfo{volume}{104}}, \bibinfo{pages}{L041103}
  (\bibinfo{year}{2021}).

\bibitem{Guo2022}
\bibinfo{author}{Guo, C.} \emph{et~al.}
\newblock \bibinfo{title}{{Switchable chiral transport in charge-ordered kagome
  metal CsV$_3$Sb$_5$}}.
\newblock \emph{\bibinfo{journal}{Nature}} \textbf{\bibinfo{volume}{611}},
  \bibinfo{pages}{461--466} (\bibinfo{year}{2022}).

\bibitem{Zheng2022}
\bibinfo{author}{Zheng, L.} \emph{et~al.}
\newblock \bibinfo{title}{{Emergent charge order in pressurized kagome
  superconductor CsV$_3$Sb$_5$}}.
\newblock \emph{\bibinfo{journal}{Nature}} \textbf{\bibinfo{volume}{611}},
  \bibinfo{pages}{682--687} (\bibinfo{year}{2022}).

\bibitem{Chen2021}
\bibinfo{author}{Chen, H.} \emph{et~al.}
\newblock \bibinfo{title}{{Roton pair density wave in a strong-coupling kagome
  superconductor}}.
\newblock \emph{\bibinfo{journal}{Nature}} \textbf{\bibinfo{volume}{599}},
  \bibinfo{pages}{222--228} (\bibinfo{year}{2021}).

\bibitem{Tazai2023}
\bibinfo{author}{Tazai, R.}, \bibinfo{author}{Yamakawa, Y.} \&
  \bibinfo{author}{Kontani, H.}
\newblock \bibinfo{title}{{Charge-loop current order and Z$_3$ nematicity
  mediated by bond order fluctuations in kagome metals}}.
\newblock \emph{\bibinfo{journal}{Nature Communications}}
  \textbf{\bibinfo{volume}{14}}, \bibinfo{pages}{7845} (\bibinfo{year}{2023}).

\bibitem{Xu2022}
\bibinfo{author}{Xu, Y.} \emph{et~al.}
\newblock \bibinfo{title}{{Three-state nematicity and magneto-optical Kerr
  effect in the charge density waves in kagome superconductors}}.
\newblock \emph{\bibinfo{journal}{Nature Physics}}
  \textbf{\bibinfo{volume}{18}}, \bibinfo{pages}{1470--1475}
  (\bibinfo{year}{2022}).

\bibitem{Nie2022}
\bibinfo{author}{Nie, L.} \emph{et~al.}
\newblock \bibinfo{title}{{Charge-density-wave-driven electronic nematicity in
  a kagome superconductor}}.
\newblock \emph{\bibinfo{journal}{Nature}} \textbf{\bibinfo{volume}{604}},
  \bibinfo{pages}{59--64} (\bibinfo{year}{2022}).

\bibitem{Li2022}
\bibinfo{author}{Li, H.} \emph{et~al.}
\newblock \bibinfo{title}{{Rotation symmetry breaking in the normal state of a
  kagome superconductor KV$_3$Sb$_5$}}.
\newblock \emph{\bibinfo{journal}{Nature Physics}}
  \textbf{\bibinfo{volume}{18}}, \bibinfo{pages}{265--270}
  (\bibinfo{year}{2022}).

\bibitem{Jin2022}
\bibinfo{author}{Jin, J.-T.}, \bibinfo{author}{Jiang, K.},
  \bibinfo{author}{Yao, H.} \& \bibinfo{author}{Zhou, Y.}
\newblock \bibinfo{title}{{Interplay between Pair Density Wave and a Nested
  Fermi Surface}}.
\newblock \emph{\bibinfo{journal}{Physical Review Letters}}
  \textbf{\bibinfo{volume}{129}}, \bibinfo{pages}{167001}
  (\bibinfo{year}{2022}).

\bibitem{Ge2025}
\bibinfo{author}{Ge, J.} \emph{et~al.}
\newblock \bibinfo{title}{{Nonreciprocal superconducting critical currents with
  normal state field trainability in kagome superconductor CsV$_3$Sb$_5$}}.
\newblock \emph{\bibinfo{journal}{arXiv}} \textbf{\bibinfo{volume}{2506}},
  \bibinfo{pages}{04601} (\bibinfo{year}{2025}).
\newblock \eprint{2506.04601}.

\bibitem{Hu2022}
\bibinfo{author}{Hu, Y.} \emph{et~al.}
\newblock \bibinfo{title}{{Topological surface states and flat bands in the
  kagome superconductor CsV$_3$Sb$_5$}}.
\newblock \emph{\bibinfo{journal}{Science Bulletin}}
  \textbf{\bibinfo{volume}{67}}, \bibinfo{pages}{495--500}
  (\bibinfo{year}{2022}).

\bibitem{Hu2022a}
\bibinfo{author}{Hu, Y.} \emph{et~al.}
\newblock \bibinfo{title}{{Rich nature of Van Hove singularities in Kagome
  superconductor CsV$_3$Sb$_5$}}.
\newblock \emph{\bibinfo{journal}{Nature Communications}}
  \textbf{\bibinfo{volume}{13}}, \bibinfo{pages}{2220} (\bibinfo{year}{2022}).

\bibitem{Liu2024}
\bibinfo{author}{Liu, Y.} \emph{et~al.}
\newblock \bibinfo{title}{{Superconductivity under pressure in a chromium-based
  kagome metal}}.
\newblock \emph{\bibinfo{journal}{Nature}} \textbf{\bibinfo{volume}{632}},
  \bibinfo{pages}{1032--1037} (\bibinfo{year}{2024}).

\bibitem{Xie2025}
\bibinfo{author}{Xie, F.} \emph{et~al.}
\newblock \bibinfo{title}{{Electron correlations in the kagome flat band metal
  CsCr$_3$Sb$_5$}}.
\newblock \emph{\bibinfo{journal}{Physical Review Research}}
  \textbf{\bibinfo{volume}{7}}, \bibinfo{pages}{L022061}
  (\bibinfo{year}{2025}).

\bibitem{Wu2025}
\bibinfo{author}{Wu, S.} \emph{et~al.}
\newblock \bibinfo{title}{{Flat-band enhanced antiferromagnetic fluctuations
  and superconductivity in pressurized CsCr$_3$Sb$_5$}}.
\newblock \emph{\bibinfo{journal}{Nature Communications}}
  \textbf{\bibinfo{volume}{16}}, \bibinfo{pages}{1375} (\bibinfo{year}{2025}).

\bibitem{Li2025}
\bibinfo{author}{Li, Y.} \emph{et~al.}
\newblock \bibinfo{title}{{Electron correlation and incipient flat bands in the
  Kagome superconductor CsCr$_3$Sb$_5$}}.
\newblock \emph{\bibinfo{journal}{Nature Communications}}
  \textbf{\bibinfo{volume}{16}}, \bibinfo{pages}{3229} (\bibinfo{year}{2025}).

\bibitem{Wang2025}
\bibinfo{author}{Wang, Z.} \emph{et~al.}
\newblock \bibinfo{title}{{Spin excitations and flat electronic bands in a
  Cr-based kagome superconductor}}.
\newblock \emph{\bibinfo{journal}{Nature Communications}}
  \textbf{\bibinfo{volume}{16}}, \bibinfo{pages}{7573} (\bibinfo{year}{2025}).

\bibitem{Peng2024}
\bibinfo{author}{Peng, S.} \emph{et~al.}
\newblock \bibinfo{title}{{Flat bands and distinct density wave orders in
  correlated Kagome superconductor CsCr$_3$Sb$_5$}}.
\newblock \emph{\bibinfo{journal}{arXiv}} \textbf{\bibinfo{volume}{2406}},
  \bibinfo{pages}{17769} (\bibinfo{year}{2024}).
\newblock \eprint{2406.17769}.

\bibitem{Meta2024}
\bibinfo{author}{Liu, L.} \emph{et~al.}
\newblock \bibinfo{title}{{Charge Density Wave Coexisting with Amplified
  Nematicity in the Correlated Kagome Metal CsCr$_3$Sb$_5$}}.
\newblock \emph{\bibinfo{journal}{Arxiv}} \textbf{\bibinfo{volume}{2411}},
  \bibinfo{pages}{1--18} (\bibinfo{year}{2024}).
\newblock \eprint{2411.06778}.

\bibitem{Kato2023}
\bibinfo{author}{Kato, T.} \emph{et~al.}
\newblock \bibinfo{title}{{Surface-termination-dependent electronic states in
  kagome superconductors AV$_3$Sb$_5$(A=K, Rb, Cs) studied by micro-ARPES}}.
\newblock \emph{\bibinfo{journal}{Physical Review B}}
  \textbf{\bibinfo{volume}{107}}, \bibinfo{pages}{245143}
  (\bibinfo{year}{2023}).

\bibitem{Kiesel2013a}
\bibinfo{author}{Kiesel, M.~L.}, \bibinfo{author}{Platt, C.} \&
  \bibinfo{author}{Thomale, R.}
\newblock \bibinfo{title}{{Unconventional Fermi Surface Instabilities in the
  Kagome Hubbard Model}}.
\newblock \emph{\bibinfo{journal}{Physical Review Letters}}
  \textbf{\bibinfo{volume}{110}}, \bibinfo{pages}{126405}
  (\bibinfo{year}{2013}).

\bibitem{Zhang2008a}
\bibinfo{author}{Zhang, W.} \emph{et~al.}
\newblock \bibinfo{title}{{High Energy Dispersion Relations for the High Temperature Bi$_2$Sr$_2$CaCu$_2$O$_8$ superconductor from Laser-Based Angle-Resolved Photoemission Spectroscopy}}.
\newblock \emph{\bibinfo{journal}{Physical Review Letters}}
  \textbf{\bibinfo{volume}{101}}, \bibinfo{pages}{017002}
  (\bibinfo{year}{2008}).

\bibitem{Chang2024}
\bibinfo{author}{Chang, M.-H.} \emph{et~al.}
\newblock \bibinfo{title}{{Dispersion kinks from electronic correlations in an
  unconventional iron-based superconductor}}.
\newblock \emph{\bibinfo{journal}{Nature Communications}}
  \textbf{\bibinfo{volume}{15}}, \bibinfo{pages}{9958} (\bibinfo{year}{2024}).

\bibitem{Krsnik2025}
\bibinfo{author}{Krsnik, J.} \& \bibinfo{author}{Held, K.}
\newblock \bibinfo{title}{{Local correlations necessitate waterfalls as a
  connection between quasiparticle band and developing Hubbard bands}}.
\newblock \emph{\bibinfo{journal}{Nature Communications}}
  \textbf{\bibinfo{volume}{16}}, \bibinfo{pages}{255} (\bibinfo{year}{2025}).

\bibitem{Zhong2023a}
\bibinfo{author}{Zhong, Y.} \emph{et~al.}
\newblock \bibinfo{title}{{Testing electron–phonon coupling for the
  superconductivity in kagome metal CsV$_3$Sb$_5$}}.
\newblock \emph{\bibinfo{journal}{Nature Communications}}
  \textbf{\bibinfo{volume}{14}}, \bibinfo{pages}{1945} (\bibinfo{year}{2023}).

\bibitem{Zhao2021b}
\bibinfo{author}{Zhao, H.} \emph{et~al.}
\newblock \bibinfo{title}{{Cascade of correlated electron states in the kagome
  superconductor CsV$_3$Sb$_5$}}.
\newblock \emph{\bibinfo{journal}{Nature}} \textbf{\bibinfo{volume}{599}},
  \bibinfo{pages}{216--221} (\bibinfo{year}{2021}).

\bibitem{Xu2025}
\bibinfo{author}{Xu, C.} \emph{et~al.}
\newblock \bibinfo{title}{{Altermagnetic ground state in distorted Kagome metal
  CsCr$_3$Sb$_5$}}.
\newblock \emph{\bibinfo{journal}{Nature Communications}}
  \textbf{\bibinfo{volume}{16}}, \bibinfo{pages}{3114} (\bibinfo{year}{2025}).

\bibitem{Han2023}
\bibinfo{author}{Han, S.} \emph{et~al.}
\newblock \bibinfo{title}{{Orbital‐Hybridization‐Driven Charge Density Wave
  Transition in CsV$_3$Sb$_5$ Kagome Superconductor}}.
\newblock \emph{\bibinfo{journal}{Advanced Materials}}
  \textbf{\bibinfo{volume}{35}} (\bibinfo{year}{2023}).

\bibitem{Huang2025a}
\bibinfo{author}{Huang, Z.} \emph{et~al.}
\newblock \bibinfo{title}{{Controlling an altermagnetic spin density wave in
  the kagome magnet CsCr$_3$Sb$_5$}}.
\newblock \emph{\bibinfo{journal}{arXiv}} \bibinfo{pages}{2510.05018}
  (\bibinfo{year}{2025}).

\bibitem{Cheng2025}
\bibinfo{author}{Cheng, S.} \emph{et~al.}
\newblock \bibinfo{title}{{Frieze charge-stripes in a correlated kagome
  superconductor CsCr$_3$Sb$_5$}}.
\newblock \emph{\bibinfo{journal}{arXiv}} \bibinfo{pages}{2510.06168}
  (\bibinfo{year}{2025}).

\bibitem{Horcas2007}
\bibinfo{author}{Horcas, I.} \emph{et~al.}
\newblock \bibinfo{title}{{WSXM : A software for scanning probe microscopy and
  a tool for nanotechnology}}.
\newblock \emph{\bibinfo{journal}{Review of Scientific Instruments}}
  \textbf{\bibinfo{volume}{78}}, \bibinfo{pages}{013705}
  (\bibinfo{year}{2007}).

\bibitem{Perdew1996}
\bibinfo{author}{Perdew, J.~P.}, \bibinfo{author}{Burke, K.} \&
  \bibinfo{author}{Ernzerhof, M.}
\newblock \bibinfo{title}{Generalized gradient approximation made simple}.
\newblock \emph{\bibinfo{journal}{Phys. Rev. Lett.}}
  \textbf{\bibinfo{volume}{77}}, \bibinfo{pages}{3865--3868}
  (\bibinfo{year}{1996}).

\bibitem{Kresse1996}
\bibinfo{author}{Kresse, G.} \& \bibinfo{author}{Furthm\"uller, J.}
\newblock \bibinfo{title}{Efficient iterative schemes for ab initio
  total-energy calculations using a plane-wave basis set}.
\newblock \emph{\bibinfo{journal}{Phys. Rev. B}} \textbf{\bibinfo{volume}{54}},
  \bibinfo{pages}{11169--11186} (\bibinfo{year}{1996}).



\end{thebibliography}

\newpage 

\newpage 

\newpage

\newpage 

\newpage

\end{document}